\documentstyle[11pt,baaa-eng,twoside,epsf,psfig]{article}
\begin{document}
\vskip 1.0cm
\markboth{J. P\'erez et al.}{}


\parindent 0pt{WORK IN PROGRESS }
\vskip 0.3cm
\title{Effects of collisions and interactions on star formation in
galaxy pairs in the field.} 
\author{Josefa P\'erez}
\affil{FCAGLP-IALP - IAFE, Buenos Aires, Argentina, jperez@fcaglp.unlp.edu.ar}
\author{Patricia Tissera}
\affil{IAFE, Buenos Aires, Argentina, patricia@iafe.uba.ar}
\author{Diego G. Lambas}
\affil{IATE, C\'ordoba, Argentina, dgl@oac.uncor.edu}

\begin{abstract}
By using cosmological simulations, we studied the effects of galaxy interactions
on the star formation activity in the local 
Universe. We selected galaxy pairs from the 3D galaxy distribution according
to a proximity criterion. The 2D galaxy catalog was constructed by projecting the 3D total galaxy
distribution and then selecting projected galaxy pairs.  
The analysis of the 3D galaxy pair catalog showed that an enhancement of the star formation activity
 can be statistically correlated with proximity.
 The projected galaxy pairs exhibited a similar trend with projected
distances and relative radial velocities. However, the star formation enhancement signal is diminished with respect to
that of the 3D galaxy pair catalog owing to projection effects and spurious galaxy pairs.
Overall, we found that hierarchical scenarios reproduced the observational dependence of star formation activity
in pairs on orbital parameters and environment. We also found that geometrical effects due to projection modify
the trends more severely than those introduced by spurious pairs. 

\end{abstract}

\section{Introduction}
Observations in the local Universe as well as at high redshift show
that galaxy interactions can trigger star formation activity independently of
environment (Lambas et al. 2003; Alonso et al. 2004).
 A possible theoretical explanation to this fact
 is associated to the dynamical stability of the systems.
Numerical simulations showed that interactions between systems
without bulge or with a small one can develop  tidal instabilities
which produce gas inflows into the central region of the systems, 
triggering starbursts (e.g. Barnes \& Hernquist 1996; Tissera 2000).

Barton et al (2000) analyzed a sample of about 250 pairs of galaxies
determining that interactions could be correlated with  an enhancement of
star formation activity. 
From 2dFGRS survey  (Colles et al 2001), Lambas et al (2003) and Alonso et al. (2004) built 
up catalogs of galaxy pairs 
in different environment, selecting them
by applying both velocity ($\Delta V \leq 350$ km $s^{-1}$) 
and projected ($r_{p}\leq 100 $ kpc) separation criteria.
These authors found  a clear correlation between interactions and the star
formation activity. 

In this work we intend to test if hierarchical  scenarios for galaxy formation 
can reproduce these observations and how projection effects might distort the real signals.

\section{Galaxy pairs in numerical simulations: Analysis and Results}

We analyze numerical simulations consistent with the {\it concordance} Cold Dark Matter
model: $\Omega_m=0.3$,  $\Lambda=0.7$ and $H_{0}=70 {\rm km s^{-1} Mpc^{-1}}$
which includes star formation and chemical enrichment.
In order to identify galactic systems from the simulations, we proceeded following
 steps. First, we identified the virialized structures by using 
the percolation method developed by Davis et al (1985), 
called {\it friends-of-friend} (fof), which
selected structures contained within a density contour defined by 
a {\it linking length} parameter. After this process, 
we defined spherical
regions of 0.5 Mpc of radio centered at each virialized system. 
Finally, within each of these regions, 
the  substructure was identified by  using a smaller 
{\it linking length} parameter. 
This procedure allowed us to select systems from $5\times 10^8$ to $10^{13}{\rm M_odot h^{-1}}$ total mass.


\vskip .1cm

After the identification of galactic systems from the simulations, we analyzed the 
physical and chemical properties of the gas and stellar components  in
 each galactic object by calculating averages over the particles within the radius that encloses $83\%$ of
the baryonic mass of the systems.
The star formation activity is quantified by the estimation of the birth rate parameter:
$b=sfr/<sfr>$,
defined as the ratio between  the present star formation rate and 
its mean value 
over whole history of the galaxy (Kennicutt et al. 1998).

\vskip .1cm
The 3D galaxy pair catalog was constructed by selecting  galaxies closer
than $r \approx 200$ kpc. Following Lambas et al. (2003) a control sample was
also defined by galaxies without a close companion. The comparison between the
star formation activity in both catalogs unveils the effects of having a close
neihbor.

The projection of the galactic systems onto randomly chosen directions allowed us
to mimic observations and to construct a 2D galaxy pair sample by applying the same
observational criteria chosen by Lambas et al. (2003). We required galaxies to be closer in 
projected distance ($r_p < 100$ kpc ) and relative velocity ($\Delta V < 350 {\rm km/s}$ in order to be included in the 2D galaxy pair catalog.
A corresponding control sample was also defined for the projected pair catalog.

\vskip .1cm
In the 3D galaxy pair catalog, we found a clear trend for an enhancement of the
star formation activity for close pairs with respect to galaxies without a close companion.
 We estimated a relative distance threshold of $\approx 50 {\rm kpc h^{-1}}$
for the star formation activity to be statistically important.
A weak trend for lower velocity encounters to
trigger stronger star formation activity was also detected.

\vskip .1cm
The 2D simulated galaxy-pair  catalog showed comparable trends  for the star formation
activity to be enhanced for small relative projected distances and relative radial velocities.
We also obtained similar dependence of the star formation activity in pairs on environment.

The 2D catalogs not only allowed us to compare the simulations with observations in a more
reliable fashion, but also to  evaluate the effects introduced by spurious pairs and projection.
We found that  many galaxies in the 2D catalog appear as pairs when
in fact their 3D relative separation is larger than the cut-off value
adopted as criteria to define pairs. These spurious systems,
with arbitrary values of star formation rate, produced a distortion in the observed trends
between star formation activity and the orbital parameters.
As expected the effects of spurious is more important for larger projected separations.
We estimated that  $\sim 29 \%$ of pairs within  $r_{p} \leq 100 {\rm kpc h^{-1}}$
are spurious while this contamination diminishes to  $16\%$  within  $r_{p} \leq 35 {\rm kpc h^{-1}}$.
Nevertheless, as discussed by Perez et al. (2005), the projection of galaxies itself introduces the
largest effects since the elimination of spurious pairs does not allow the recovering of the original signal.

\section{Conclusions.}

Comparing the results from 3D and 2D simulated galaxy pair catalogs, we conclude that:

\vskip .1cm
\hskip 1cm $\bullet$ there is a correlation between the star formation activity of a galaxy and the 3D
distance to its closest neighbor, which have to be, on averaged, at $ r < 50 {\rm kpc/h}$ to show
important star formation enhancement with respect to isolated systems. 

\vskip .1cm
\hskip 1cm $\bullet$ the properties of galaxy pairs in the 2D simulated catalog reproduced
the observations trends with orbital parameters and environment obtained from Lambas et al. (2003).

\vskip .1cm
\hskip 1cm $\bullet$ spurious pairs in the 2D galaxy pair catalog are more imporant at larger separation and their
effects are less severe than those introduced by the projection of the 3D galaxy distribution itself.

\section{Acknowledgments}
This work was partially supported by Fundaci\'on Antorchas and Consejo Nacional de Investigaciones
Cient\'\i ficas y T\'ecnicas.

\end{document}